\documentclass[aip,jcp,reprint,noshowkeys,superscriptaddress]{revtex4-2}
\usepackage{graphicx,dcolumn,bm,xcolor,microtype,multirow,amscd,amsmath,amssymb,amsfonts,physics,wrapfig,txfonts,siunitx,xspace}
\usepackage[version=4]{mhchem}

\usepackage[utf8]{inputenc}
\usepackage[T1]{fontenc}
\usepackage[normalem]{ulem}
\usepackage{tikz}
\usepackage{bbold}

\usepackage[
	colorlinks=true,
	citecolor=blue,
	breaklinks=true,
	urlcolor=purple,
	]{hyperref}
\usepackage{orcidlink}

\usepackage{mathtools}
\usepackage{comment}
\usepackage[nolist]{acronym}

\usepackage{cancel}

\usetikzlibrary{positioning}

\newcommand{\hH}{\Hat{H}}
\newcommand{\hh}{\Hat{h}}
\newcommand{\hV}{\Hat{V}}
\newcommand{\hL}{\Hat{L}}
\newcommand{\hO}{\Hat{O}}

\newcommand{\bO}{\boldsymbol{0}}
\newcommand{\bI}{\boldsymbol{1}}

\newcommand{\bA}{\boldsymbol{A}}
\newcommand{\bB}{\boldsymbol{B}}
\newcommand{\bC}{\boldsymbol{C}}
\newcommand{\bM}{\boldsymbol{M}}
\newcommand{\bL}{\boldsymbol{L}}
\newcommand{\bS}{\boldsymbol{S}}
\newcommand{\bR}{\boldsymbol{R}}
\newcommand{\bT}{\boldsymbol{T}}
\newcommand{\bX}{\boldsymbol{X}}
\newcommand{\brX}{\bX}
\newcommand{\blX}{\Tilde{\bX}}
\newcommand{\bY}{\boldsymbol{Y}}
\newcommand{\brY}{\bY}
\newcommand{\blY}{\Tilde{\bY}}
\newcommand{\bOm}{\boldsymbol{\Omega}}
\newcommand{\bDel}{\boldsymbol{\Delta}}

\newcommand{\bb}{\boldsymbol{b}}

\newcommand{\SupMat}{\textcolor{blue}{Supplementary Materials}\xspace}
\newcommand{\T}[1]{#1^{\intercal}}

\newcommand{\opa}[1]{\Hat{a}_{#1}}
\newcommand{\opad}[1]{\Hat{a}^{\dagger}_{#1}}
\newcommand{\duala}[1]{\Tilde{a}_{#1}}
\newcommand{\dualad}[1]{\Tilde{a}^{\dagger}_{#1}}

\newcommand{\opbd}[1]{\Hat{b}^{\dagger}_{#1}}
\newcommand{\dualb}[1]{\Tilde{b}_{#1}}

\newcommand{\opQ}[1]{\Hat{Q}_{#1}}
\newcommand{\opQd}[1]{\Hat{Q}^{\dagger}_{#1}}
\newcommand{\dualQ}[1]{\Tilde{Q}_{#1}}
\newcommand{\dualQd}[1]{\Tilde{Q}^{\dagger}_{#1}}

\newcommand{\Ne}{N}

\newcommand{\rphi}[1]{{\phi_{#1}}}
\newcommand{\lphi}[1]{{\tilde{\phi}_{#1}}}
\newcommand{\rPsi}[1]{{\Psi_{#1}}}
\newcommand{\lPsi}[1]{{\tilde{\Psi}_{#1}}}
\newcommand{\rPhi}[1]{{\Phi_{#1}}}
\newcommand{\lPhi}[1]{{\tilde{\Phi}_{#1}}}
\newcommand{\rX}[2]{{X_{#1}^{#2}}}
\newcommand{\lX}[2]{{\tilde{X}_{#1}^{#2}}}
\newcommand{\rY}[2]{{Y_{#1}^{#2}}}
\newcommand{\lY}[2]{{\tilde{Y}_{#1}^{#2}}}

\newcommand{\br}[1]{{\boldsymbol{r}_{#1}}}

\usepackage{csquotes}

\begin{document}

\newcommand{\LCPQ}{Laboratoire de Chimie et Physique Quantiques (UMR 5626), Universit\'e de Toulouse, CNRS, UPS, France}
\newcommand{\LPCT}{Laboratoire de Physique et Chimie Th\'eoriques (UMR 7019), Universit\'e de Lorraine, CNRS, F-54000 Nancy, France}
\newcommand{\EPFL}{Institute of Physics, \'Ecole Polytechnique F\'ed\'erale de Lausanne (EPFL), CH-1015 Lausanne, Switzerland}
\newcommand{\LPT}{Laboratoire de Chimie Théorique, Sorbonne Université and CNRS, F-75005 Paris, France}

\author{Abdallah Ammar \orcidlink{0009-0009-2774-0520}}
	\email{abdallah.ammar@univ-lorraine.fr}
\affiliation{\LPCT}
\author{Enzo Monino \orcidlink{0009-0001-6907-9058}}
	\email{enzo.monino@epfl.ch}
\affiliation{\EPFL}
\author{Anthony Scemama \orcidlink{0000-0003-4955-7136}}
	\email{scemama@irsamc.ups-tlse.fr}
\affiliation{\LCPQ}
\author{Emmanuel Giner \orcidlink{0000-0002-6206-1103}}
	\email{eginer@lct.jussieu.fr}
\affiliation{\LPT}
\author{Pierre-Fran\c{c}ois Loos \orcidlink{0000-0003-0598-7425}}
	\email{loos@irsamc.ups-tlse.fr}
\affiliation{\LCPQ}

\newcommand{\Ort}{orthogonal}
\newcommand{\BiO}{bi-orthogonal}

\title{Transcorrelated Random-Phase Approximation}

\begin{abstract}
We extend the random-phase approximation (RPA) to the non-Hermitian transcorrelated (TC) Hamiltonian, which explicitly includes three-body interactions generated by a Jastrow correlation factor. 
We consider both the direct RPA (dRPA) and RPA with exchange (RPAx). 
We apply the resulting TC-dRPA and TC-RPAx methods to calculate ground-state correlation energies and vertical excitation energies for atoms (\ce{He} and \ce{Ne}) and small molecules (\ce{H2O}, \ce{NH3}, \ce{CH4}, and \ce{H2CO}). 
For ground-state correlation energies, the TC treatment substantially improves accuracy and accelerates basis set convergence, reducing errors by nearly an order of magnitude relative to conventional RPA calculations. 
By contrast, it yields only marginal improvements in vertical excitation energies. 
We attribute this limited effect to the ground-state optimization of the Jastrow factor, which does not adequately capture the distinct electronic character of excited states. 
These results establish TC-RPA as an accurate and computationally efficient approach to ground-state energetics, while highlighting the need for state-specific Jastrow optimization to achieve reliable descriptions of excited states.
\end{abstract}

\maketitle

\begin{acronym}
  \acro{FCI}{full configuration interaction}
  \acro{FROGG}{frozen Gaussian geminal}
  \acro{CIS}{configuration interaction singles}
  \acro{CI}{configuration interaction}
  \acro{SCI}{selected configuration interaction}
  \acro{CC}{coupled-cluster}
  \acro{EN}{Epstein-Nesbet}
  \acro{QMC}{quantum Monte Carlo}
  \acro{AO}{atomic orbital}
  \acro{MO}{molecular orbital}
  \acro{HF}{Hartree-Fock}
  \acro{TDHF}{time-dependent Hartree-Fock}
  \acro{BHF}{Brueckner-Hartree-Fock}
  \acro{CAS}{complete active space}
  \acro{VMC}{variational Monte Carlo}
  \acro{DMC}{diffusion Monte Carlo}
  \acro{TC}{transcorrelated}
  \acro{CASSCF}{complete active space self consistent field}
  \acro{fc}{frozen core}
  \acro{FCIQMC}{full configuration interaction quantum Monte Carlo}
  \acro{SCF}{self-consistent field}
  \acro{RHF}{restricted Hartree-Fock}
  \acro{UHF}{unrestricted Hartree-Fock}
  \acro{ROHF}{restricted open-shell Hartree-Fock}
  \acro{DIIS}{direct inversion of the iterative subspace}
  \acro{LS}{level-shift}
  \acro{MP}{M{\o}ller-Plesset}
  \acro{PT}{perturbation theory}
  \acro{MPPT2}[MP2]{M{\o}ller-Plesset perturbation theory at second order}
  \acro{BiO}{bi-orthogonal}
  \acro{MPS}{matrix product state}
  \acro{MBPT2}[MBPT(2)]{many-body perturbation theory at second order}
  \acro{LCCSD}{linearized coupled-cluster singles and doubles}
  \acro{DMRG}{density matrix renormalization group}
  \acro{DFT}{density-functional theory}
  \acro{TDDFT}{time-dependent density-functional theory}
  \acro{WFT}{wave function theory}
  \acro{Var}{Variational}
  \acro{CBS}{complete basis set}
  \acro{IP}{ionization potential}
  \acro{AE}{atomization energy}
  \acro{TDA}{Tamm-Dancoff approximation}
  \acro{RPA}{random-phase approximation}
  \acro{SRPA}{second random-phase approximation}
  \acro{dRPA}{direct random-phase approximation}
  \acro{RPAx}{random-phase approximation with exchange}
  \acro{SCRPA}{self-consistent random-phase approximation}
  \acro{QRPA}{quasi-particle random-phase approximation}
  \acro{EOM}{equation-of-motion}
  \acro{BQM}{biorthogonal quantum mechanics}
  \acro{QBA}{quasiboson approximation}
  \acro{ERI}{electron repulsion integrals}
  \acro{TDDM}{time-dependent density-matrix}
  \acro{STDDM}{small-amplitude time-dependent density-matrix}
  \acro{DMM}{density matrix methods}
\end{acronym}

\section{Introduction}

The difficulty of electronic structure calculations is not only that the many-electron Schr\"odinger equation is hard to solve, but also that its exact solution has features that are poorly matched to standard orbital expansions. 
In particular, the electronic wavefunction has a cusp at the electron-electron coalescence points, \cite{Kato_1951,Kato_1957,Myers_1991,Morgan_1993} a short-range nonanalyticity that finite Gaussian basis sets recover only slowly. \cite{Schwartz_1962,Hill_1985,Kutzelnigg_1992,Hattig_2012,Kong_2012}
Even a formally exact treatment of correlation within a given orbital space can therefore retain sizeable basis set incompleteness errors, motivating approaches that build short-range correlation directly into the theory.

This issue is especially relevant for the \ac{RPA}, \cite{Bohm_1951,Pines_1952,Bohm_1953} which provides a compact description of long-range correlation and collective electronic fluctuations, but converges slowly with respect to the one-electron basis set. \cite{Eshuis_2012_Basis,Hehn_2013,Hehn_2015,Traore_2023} 
At the same time, \ac{RPA} occupies a distinctive position in quantum chemistry because of its close formal connections to coupled-cluster theory, \cite{Freeman_1977,Scuseria_2008,Scuseria_2013,Peng_2013,Lange_2018,Berkelbach_2018,Rishi_2020,Monsebaiz_2022,Tolle_2023,Tolle_2026} \ac{DFT},\cite{Furche_2008,Jansen_2010,Angyan_2011,Eshuis_2012,Trushin_2021,Trushin_2025} and one-body Green's function methods. \cite{Hedin_1965,Aryasetiawan_1998,Onida_2002,Reining_2017,Golze_2019,Marie_2024_GW} 
This combination of theoretical appeal and practical basis set sensitivity makes \ac{RPA} an attractive target for explicitly correlated or Hamiltonian-transformed approaches.

The \ac{RPA} was originally developed in plasma, condensed-matter, and nuclear physics, \cite{Bohm_1951,Pines_1952,Bohm_1953,Sawada_1957,Brout_1957,Wentzel_1957,Gell-Mann_1957,Hubbard_1958,Nozieres_1958,Thouless_1960,Thouless_1961,Brown_1961} where it became a foundational many-body technique. In quantum chemistry, it has been used both for atomic and molecular excited-state calculations. \cite{Ball_1964,Dunning_1967,Dunning_1968,Linderberg_1970,Shibuya_1970_Application,Ostlund_1971,Jorgensen_1972,Ho_1972,Tanimoto_1972,Jordan_1973,Ren_2012,Chen_2017,Jemai_2020,Schuck_2021}
In its conventional \ac{EOM} formulation, \ac{RPA} restricts the excitation operator to the one-hole-one-particle manifold and invokes the quasiboson approximation. 
The latter replaces the exact commutators of fermionic particle-hole operators by their expectation values over an uncorrelated \ac{HF} reference. 
Particle-hole pairs are therefore treated as approximately bosonic collective modes, yielding a simple linear eigenvalue problem. This simplification does not exactly preserve the Pauli constraints associated with the underlying fermionic operators and can lead to an overestimation of ground-state correlation effects. \cite{Singwi_1968,Jiang_2007,Forster_2022}

Many extensions of \ac{RPA} have been proposed to improve its formal structure, including perturbative corrections, \cite{Takayanagi_1988_gen,Takayanagi_1988_theo,Nishizaki_1988,Adachi_1988,Drozdz_1990,Mariano_1994,Joshi_2024} density-matrix formulations, \cite{Chatterjee_2012,Vanaggelen_2013,Pernal_2014,Pernal_2014_Intergeminal,DePrince_2016,Pernal_2018,Maradzike_2018} self-consistent \ac{RPA} variants, \cite{Kerman_1963,Hara_1964,Dang_1964,Ikeda_1965,Providencia_1965,Rowe_1968_Methods,Schuck_2021} boson-mapping approaches, \cite{Marshalek_1969,Marshalek_1970,Marshalek_1982,Marshalek_1987_Generic,Catara_1989,Sambataro_1995,Sambataro_1997,Gambacurta_2006_Solvable} and second-\ac{RPA} schemes including two-hole-two-particle excitations. \cite{Werthamer_1962,Fano_1962,Sawicki_1962,Tamura_1964,Adachi_1984,Drozdz_1986_Giant,Gambacurta_2009,Gambacurta_2010_collect,Papakonstantinou_2010,Peng_2014} Particle-particle and quasiparticle formulations have also provided useful alternatives, especially for strongly correlated systems and pairing correlations. \cite{Hogaasen_1961,Hirsch_2002,Dukelsky_2003,Scuseria_2013,Peng_2013,Aggelen_2013,Yang_2013,Zhang_2015,Berkelbach_2018} 
These developments address important limitations of the standard approximation, but they do not directly target the short-range correlation responsible for slow basis set convergence in \ac{RPA} calculations.

Several strategies have been designed to alleviate this short-range deficiency. 
Range-separated \ac{RPA} partitions the electron-electron interaction into a short-range \ac{DFT} component and a long-range \ac{RPA} component. \cite{Toulouse_2009,Toulouse_2010} 
Explicitly correlated F12 methods instead accelerate basis set convergence by introducing geminal functions that depend explicitly on the interelectronic distance. \cite{Kutzelnigg_1985,Kutzelnigg_1991,Noga_1994,Tenno_2012a,Hattig_2012,Kong_2012,Hehn_2015} 
A related density-functional route, developed by Toulouse, Giner, and co-workers, corrects finite-basis wavefunction calculations using short-range correlation functionals designed to recover the correlation missing from an incomplete orbital space.~\cite{Giner_2018,Loos_2019d,Giner_2019,Loos_2020a,Giner_2020,Traore_2023} 
These approaches all point to the same physical requirement: an efficient \ac{RPA} scheme should incorporate the electron-electron cusp, or at least its dominant short-range consequences, more directly than is possible with a finite orbital expansion alone.

In this work, we pursue this goal through \ac{TC} theory. 
In the transcorrelated approach, dynamical correlation is incorporated into the Hamiltonian by a similarity transformation generated by a Jastrow correlation factor. \cite{Hirschfelder_1963,Boys_1969,Jastrow_1955} 
The transformed Hamiltonian is non-Hermitian and contains effective two- and three-body interaction terms. 
This Hamiltonian-level treatment of short-range correlation has been successfully combined with several post-\ac{HF} methods, including second-order many-body perturbation theory, \cite{Hino_2001,Ochi_2015} (selected) configuration interaction, \cite{Zweistra_2003,Tsuneyuki_2008,Luo_2011,Ochi_2014,Giner_2021,Ammar_2022_Extension,Ammar_2022_Optimization} \ac{FCIQMC}, \cite{Luo_2018,Dobrautz_2019,Cohen_2019,Dobrautz_2022,Jeszenszki_2020,Guther_2021,Haupt_2023} coupled-cluster theory, \cite{Hino_2002,TenNo_2002,Liao_2021,Schraivogel_2021} \ac{DMRG}, \cite{Baiardi_2020,Baiardi_2022} and \ac{DFT}. \cite{Imamura_2003,Umezawa_2017}

Here, we derive the \ac{RPA} equations for the non-Hermitian \ac{TC} Hamiltonian, including the contributions induced by the explicit three-body interaction. 
This defines a \ac{TC}-\ac{RPA} framework in which short-range correlation is incorporated at the Hamiltonian level before the linear-response problem is constructed. 
We consider two variants: \ac{dRPA}, in which only the direct Coulomb contribution to the particle-hole interaction kernel is retained, and \ac{RPAx}, in which exchange is included through the corresponding antisymmetrized interaction kernel.
The resulting \ac{TC}-\ac{dRPA} and \ac{TC}-\ac{RPAx} methods are assessed for both ground-state correlation energies and vertical excitation energies. 
This dual benchmark is important because the Jastrow factor used in the \ac{TC} Hamiltonian is optimized for the ground state: one may therefore expect the largest benefits for ground-state energetics, while excited-state properties provide a more stringent test of transferability.

The remainder of this article is organized as follows. 
In Sec.~\ref{sec:RPA}, we derive the \ac{TC}-\ac{RPA} equations using the \ac{EOM} formalism, analyze the structure of the resulting non-Hermitian response problem, and present spin-adapted working equations for both variants. 
In Sec.~\ref{sec:res}, we benchmark these methods against their conventional \ac{RPA} counterparts.
We draw our conclusions in Sec.~\ref{sec:ccl}.

\section{Random-phase approximation for a non-Hermitian Hamiltonian}
\label{sec:RPA}

\subsection{Transcorrelated Hamiltonian}

The \ac{RPA} can be derived from several theoretical frameworks, including the adiabatic-connection fluctuation-dissipation theorem,\cite{Mclachlan_1964,Harris_1974,Langreth_1975,Gunnarsson_1976,Langreth_1977,Toulouse_2009,Toulouse_2010,Eshuis_2012,Mussard_2015}
coupled-cluster theory,\cite{Freeman_1977,Scuseria_2008,Jansen_2010,Lotrich_2011,Klopper_2011,Jemai_2013,Scuseria_2013,Peng_2013,Szabados_2017,Berkelbach_2018,Margocsy_2020,Rishi_2020}
many-body perturbation theory,\cite{Dahlen_2006,Ren_2012,Caruso_2013}
and \ac{TDHF} or \ac{TDDFT}.\cite{Ball_1964,Ando_1977,Oddershede_1977,Jorgensen_1975,Oddershede_1978,Zangwill_1980,Stott_1980,Olov_1990,Furche_2001,Eshuis_2012}
Here, we adopt the \ac{EOM} approach.\cite{Rowe_1966,Rowe_1968_EOM}
A detailed derivation is provided in the \SupMat.
Throughout this work, we restrict our attention to closed-shell systems. Occupied, virtual, and general spin-orbitals are denoted by $i,j,\dots$, $a,b,\dots$, and $p,q,\dots$, respectively, while the corresponding spatial orbitals are denoted by $I,J,\dots$, $A,B,\dots$, and $P,Q,\dots$.

We outline below the key steps leading to the \ac{RPA} equations for a three-body non-Hermitian Hamiltonian of the form
\begin{equation}
        \hH
        =
        \hh
        + \frac{1}{2!} \sum_{i}^{\Ne} \sum_{j \neq i}^{\Ne} \hV(\br{i},\br{j})
        + \frac{1}{3!} \sum_{i}^{\Ne} \sum_{j \neq i}^{\Ne} \sum_{k \notin \{i,j\}}^{\Ne} \hL (\br{i},\br{j},\br{k}).
\label{eq:Hdef}
\end{equation}
This is the standard structure of the \ac{TC} Hamiltonian, where $\hh$, $\hV$, and $\hL$ denote the one-, two-, and three-body contributions, respectively.
The non-Hermitian character of $\hV$, and therefore of $\hH$, originates from the gradient terms generated by the similarity transformation.
Explicit definitions and detailed expressions for these operators are given in Sec.~II.A of Ref.~\onlinecite{Ammar_2024}.

\subsection{Biorthogonal formalism}

Because $\hH$ is non-Hermitian, we formulate the theory in a biorthogonal one-particle basis.\cite{Moshinsky_1971,Brody_2014}
Let $\{\rphi{p}\}$ denote a set of right spin-orbitals that is complete but not necessarily orthonormal.
The associated creation and annihilation operators then satisfy
\begin{equation}
        \acomm{ \opad{p} }{ \opa{q} } = \braket*{\phi_p}{\phi_q} = S_{pq} \neq \delta_{pq},
\end{equation}
where $\acomm{\cdot}{\cdot}$ denotes the anticommutator.
To recover canonical anticommutation relations, we introduce the dual basis $\{\lphi{p}\}$,
\begin{equation}
	\lphi{p} = \sum_q \qty(\bS^{-1})_{pq} \rphi{q},
\end{equation}
such that $\braket*{\lphi{p}}{\rphi{q}} = \delta_{pq}$.
Using the corresponding dual annihilation operators, one obtains
\begin{align}
        \acomm{\duala{p} }{ \opad{q} } &= \delta_{pq},
        &
        \acomm{\opad{p} }{ \opad{q} } &= 0,
        &
        \acomm{\duala{p} }{ \duala{q} } &= 0.
        \label{eq:anticomm_dual}
\end{align}

In this biorthogonal representation, the second-quantized Hamiltonian reads
\begin{equation}
\begin{split}
	\hH
	& = \sum_{p,q} h_q^p \, \opad{p} \duala{q}
	+ \frac{1}{2!} \sum_{p_1,p_2} \sum_{q_1,q_2}
	V_{q_1 q_2}^{p_1 p_2} \, \opad{p_1} \opad{p_2} \duala{q_2} \duala{q_1} \\
	& + \frac{1}{3!} \sum_{p_1,p_2,p_3} \sum_{q_1,q_2,q_3}
	L_{q_1 q_2 q_3}^{p_1 p_2 p_3}
	\opad{p_1} \opad{p_2} \opad{p_3} \duala{q_3} \duala{q_2} \duala{q_1} \\
	& =
	\sum_{p,q} h_q^p \, \opad{p} \duala{q}
	+ \frac{1}{4} \sum_{p_1,p_2} \sum_{q_1,q_2}
	\bar{V}_{q_1 q_2}^{p_1 p_2} \, \opad{p_1} \opad{p_2} \duala{q_2} \duala{q_1} \\
	& + \frac{1}{36} \sum_{p_1,p_2,p_3} \sum_{q_1,q_2,q_3}
	\bar{L}_{q_1 q_2 q_3}^{p_1 p_2 p_3}
	\opad{p_1} \opad{p_2} \opad{p_3} \duala{q_3} \duala{q_2} \duala{q_1}.
\end{split}
\end{equation}
The one-, two-, and three-body integrals are defined as
\begin{subequations}
\begin{align}
	\label{eq:1e_integ}
	h_q^p
	&= \mel*{\lphi{p}}{\hh}{\rphi{q}}, \\
	\label{eq:2e_integ}
	V_{q_1 q_2}^{p_1 p_2}
	&= \mel*{\lphi{p_1} \lphi{p_2}}{\hV}{\rphi{q_1} \rphi{q_2}}, \\
	\label{eq:3e_integ}
	L_{q_1 q_2 q_3}^{p_1 p_2 p_3}
	&= \mel*{\lphi{p_1} \lphi{p_2} \lphi{p_3}}{\hL}{\rphi{q_1} \rphi{q_2} \rphi{q_3}}.
\end{align}
\end{subequations}
We also introduce the antisymmetrized matrix elements
\begin{subequations}
\begin{align}
	\bar{V}_{q_1 q_2}^{p_1 p_2}
	& =
	V_{q_1 q_2}^{p_1 p_2} - V_{q_2 q_1}^{p_1 p_2}, \\
	\begin{split}
		\bar{L}_{q_1 q_2 q_3}^{p_1 p_2 p_3}
		& =
		L_{q_1 q_2 q_3}^{p_1 p_2 p_3}
		- L_{q_1 q_3 q_2}^{p_1 p_2 p_3}
		+ L_{q_2 q_3 q_1}^{p_1 p_2 p_3} \\
		&
		- L_{q_2 q_1 q_3}^{p_1 p_2 p_3}
		+ L_{q_3 q_1 q_2}^{p_1 p_2 p_3}
		- L_{q_3 q_2 q_1}^{p_1 p_2 p_3}.
	\end{split}
\end{align}
\end{subequations}

Finally, we assume that the left and right spin-orbitals diagonalize the non-symmetric Fock operator,\cite{Ammar_2022_Optimization,Ammar_2022_Extension,Ammar_2023_Biorthonormal,Ammar_2023_Transcorrelated,Ammar_2024}
\begin{equation} \label{eq:TC-Fock}
	F_q^p
	= h_q^p + \sum_{i} \bar{V}_{i q}^{i p} + \frac{1}{2} \sum_{i,j} \bar{L}_{i j q}^{i j p}
	= \epsilon_p \, \delta_{pq}.
\end{equation}

\subsection{Equation-of-motion formalism}

Let $\opQd{\nu}$ and $\dualQd{\nu}$ be excitation operators that generate the right and left excited states from the corresponding \ac{RPA} ground states,\cite{Reinhard_1992}
\begin{align}
	\opQd{\nu}   \ket*{\rPsi{0}} &= \ket*{\rPsi{\nu}},
	&
	\dualQd{\nu} \ket*{\lPsi{0}} &= \ket*{\lPsi{\nu}},
\end{align}
with $\braket*{\lPsi{\mu}}{\rPsi{\nu}} = \delta_{\mu\nu}$.
The right and left eigenstates satisfy
\begin{align}
	\hH \ket*{\rPsi{\mu}} &= E_{\mu} \ket*{\rPsi{\mu}},
	&
	\hH^\dagger \ket*{\lPsi{\mu}} = E_{\mu} \ket*{\lPsi{\mu}}.
\end{align}
where the energies are assumed to be real, i.e., $E_{\mu} = E_{\mu}^*$.
The direct and dual \ac{RPA} ground states are assumed to exist and to satisfy the killing conditions
$\opQ{\nu} \ket*{\rPsi{0}} = \dualQ{\nu} \ket*{\lPsi{0}} = 0$.

As shown in the \SupMat, for an arbitrary operator $\hO$, the following two \ac{EOM} identities hold:
\begin{subequations}
\begin{align}
	\mel{\tilde{\Psi}_0}{\comm{\hO }{ \comm{ \hH }{ \opQd{\nu} } } }{\Psi_0}
	& = \Omega_{\nu} \mel{\tilde{\Psi}_0}{\comm{ \hO }{ \opQd{\nu} } }{\Psi_0}, \\
	\mel{\tilde{\Psi}_0}{\comm{ \comm{ \dualQ{\nu} }{ \hH } }{ \hO } }{\Psi_0}
	& = \Omega_{\nu} \mel{\tilde{\Psi}_0}{\comm{ \dualQ{\nu} }{ \hO } }{\Psi_0}.
\end{align}
\end{subequations}
Here, $\comm{\cdot}{\cdot}$ denotes the commutator and $\Omega_\nu = E_{\nu} - E_0$ is the excitation energy associated with the $\nu$th excited state.
The two \ac{EOM} expressions become identical only in the Hermitian limit.

The \ac{RPA} equations are obtained by using the excitation operators
\begin{subequations}
\begin{align}
	 \opQd{\nu}   &= \sum_{ia} \rX{ia}{\nu} \opad{a} \duala{i} - \sum_{ia} \rY{ia}{\nu} \opad{i} \duala{a}, \\
	 \dualQd{\nu} &= \sum_{ia} \lX{ia}{\nu} \dualad{a} \opa{i} + \sum_{ia} \lY{ia}{\nu} \dualad{i} \opa{a},
\end{align}
\end{subequations}
and by choosing the first-order variations $\hO=\delta \dualQ{\nu}$ and $\hO=\delta \opQd{\nu}$ in the first and second \ac{EOM}, respectively.
This yields the two sets of equations
\begin{subequations}
\begin{align}
	\sum_{jb} \qty[+A_{ia,jb} \rX{jb}{\nu} + B_{ia,jb} \rY{jb}{\nu}]
	&= \Omega_{\nu} \mel{\lPsi{0}}{\comm{ \opad{i} \duala{a} }{ \opQd{\nu} } }{\rPsi{0}}, \\
	\sum_{jb} \qty[-C_{ia,jb} \rX{jb}{\nu} - D_{ia,jb} \rY{jb}{\nu}]
	&= \Omega_{\nu} \mel{\lPsi{0}}{\comm{ \opad{a} \duala{i} }{ \opQd{\nu} } }{\rPsi{0}},
\end{align}
\end{subequations}
and
\begin{subequations}
\begin{align}
	\sum_{jb} \qty[+\tilde{A}_{ia,jb} \lX{jb}{\nu} - \tilde{C}_{ia,jb} \, \lY{jb}{\nu}]
	&= \Omega_{\nu} \mel{\lPsi{0}}{\comm{ \dualQ{\nu} }{ \opad{a} \duala{i} } }{\rPsi{0}}, \\
	\sum_{jb} \qty[-\tilde{B}_{ia,jb} \lX{jb}{\nu} + \tilde{D}_{ia,jb} \, \lY{jb}{\nu}]
	&= \Omega_{\nu} \mel{\lPsi{0}}{\comm{ \dualQ{\nu} }{ \opad{i} \duala{a} } }{\rPsi{0}}.
\end{align}
\end{subequations}
The matrices appearing in these equations are
\begin{subequations}
\begin{align}
	A_{ia,jb} &= +\mel{\lPsi{0}}{\comm{ \opad{i} \duala{a} }{ \comm{ \hH }{ \opad{b} \duala{j} } } }{\rPsi{0}}, \\
	B_{ia,jb} &= -\mel{\lPsi{0}}{\comm{ \opad{i} \duala{a} }{ \comm{ \hH }{ \opad{j} \duala{b} } } }{\rPsi{0}}, \\
	C_{ia,jb} &= -\mel{\lPsi{0}}{\comm{ \opad{a} \duala{i} }{ \comm{ \hH }{ \opad{b} \duala{j} } } }{\rPsi{0}}, \\
	D_{ia,jb} &= +\mel{\lPsi{0}}{\comm{ \opad{a} \duala{i} }{ \comm{ \hH }{ \opad{j} \duala{b} } } }{\rPsi{0}},
\end{align}
\end{subequations}
and
\begin{subequations}
\begin{align}
	\tilde{A}_{ia,jb} &= +\mel{\lPsi{0}}{\comm{ \comm{ \opad{j} \duala{b} }{ \hH } }{ \opad{a} \duala{i} } }{\rPsi{0}}, \\
	\tilde{B}_{ia,jb} &= -\mel{\lPsi{0}}{\comm{ \comm{ \opad{j} \duala{b} }{ \hH } }{ \opad{i} \duala{a} } }{\rPsi{0}}, \\
	\tilde{C}_{ia,jb} &= -\mel{\lPsi{0}}{\comm{ \comm{ \opad{b} \duala{j} }{ \hH } }{ \opad{a} \duala{i} } }{\rPsi{0}}, \\
	\tilde{D}_{ia,jb} &= +\mel{\lPsi{0}}{\comm{ \comm{ \opad{b} \duala{j} }{ \hH } }{ \opad{i} \duala{a} } }{\rPsi{0}}.
\end{align}
\end{subequations}

At this stage, we introduce the quasiboson approximation. \cite{Ring_2004}
The commutator between the direct particle-hole annihilation operator $\dualb{\mu} \equiv \opad{i}\duala{a}$ and the dual particle-hole creation operator $\opbd{\nu} \equiv \opad{b}\duala{j}$ is approximated as 
\begin{equation}
	\comm{\dualb{\mu}}{\opbd{\nu}}
	= \delta_{ab} \opad{i}\duala{j} - \delta_{ij} \opad{b}\duala{a}
	\approx \delta_{ij}\delta_{ab}
	= \delta_{\mu\nu}.
\end{equation}
Equivalently, the quasiboson operators are assumed to satisfy
\begin{align}
	\qty[\dualb{\mu}, \opbd{\nu}] &= \delta_{\mu \nu},
	&
	\qty[\opbd{\mu}, \opbd{\nu}] &= 0,
	&
	\qty[\dualb{\mu}, \dualb{\nu}] &= 0.
	\label{eq:qba}
\end{align}
In addition, expectation values over the correlated \ac{RPA} ground states are replaced by expectation values over the \ac{HF} Slater determinant and its dual,
$\mel*{\lPsi{0}}{\cdots}{\rPsi{0}} \rightarrow \mel*{\lPhi{0}}{\cdots}{\rPhi{0}}$.
The quasiboson approximation therefore treats fermionic particle-hole excitations as effective bosonic modes.
This approximation considerably simplifies the many-body problem, but it relaxes the Pauli constraints imposed by the underlying fermionic algebra and can lead to an artificial lowering of the ground-state energy.

After applying the quasiboson approximation (see the \SupMat for details), the amplitudes $\brX,\brY$ and $\blX,\blY$ are obtained as right and left eigenvectors of the \ac{RPA} matrix
\begin{equation}
	\bM
	=
	\begin{pmatrix}
		+\bA & +\bB \\
		-\bC & -\T{\bA}
        \end{pmatrix}.
\label{eq:Mdef}
\end{equation}
The matrix blocks are
\begin{subequations}
\begin{align}
	A_{ia,jb}
	&= \qty(\epsilon_a-\epsilon_i) \delta_{ab} \delta_{ij} + \bar{U}_{i b}^{a j}, \\
	B_{ia,jb}
	&= \bar{U}_{i j}^{a b}, \\
	C_{ia,jb}
	&= \bar{U}_{a b}^{i j},
\end{align}
\end{subequations}
where
\begin{equation}
	\bar{U}_{q_1 q_2}^{p_1 p_2}
	=
	\bar{V}_{q_1 q_2}^{p_1 p_2} + \sum_{k} \bar{L}_{k q_1 q_2}^{k p_1 p_2}.
\end{equation}
The blocks $\bB$ and $\bC$ are symmetric, whereas $\bA$ is not.
Indeed, the non-Hermiticity of $\hV$ implies
$\bar{V}_{q_1 q_2}^{p_1 p_2} \ne \bar{V}_{p_1 p_2}^{q_1 q_2} \ne \bar{V}_{p_2 p_1}^{q_2 q_1}$,
and analogous relations hold for $\bar{U}$.
Consequently, the \ac{RPA} matrix in Eq.~\eqref{eq:Mdef} contains four distinct blocks, instead of the two independent blocks that appear in conventional Hermitian \ac{RPA}.

Although the \ac{TC}-\ac{RPA} eigenvalue problem retains a structure similar to that of conventional \ac{RPA}, its individual components have a markedly different physical content. 
First, the orbital-energy differences entering the diagonal part of $\bA$ are constructed from \ac{TC}-\ac{HF} orbital energies rather than conventional \ac{HF} ones. 
Indeed, the \ac{TC} Fock operator defined in Eq.~\eqref{eq:TC-Fock} contains contractions of both the Jastrow-transformed two-electron interaction and the explicit three-electron operator. 
The resulting orbital energies can therefore be regarded as effective quantities that are already dressed by the short-range correlation encoded in the Jastrow factor. 
Second, the effective particle-hole interaction $\bar{U}$ is no longer constructed from the bare Coulomb interaction. 
Its two-electron component generates Jastrow-dressed direct and exchange-like contributions, while the contraction $\sum_k \bar{L}_{kq_1q_2}^{kp_1p_2}$ introduces an additional effective two-electron contribution of three-body origin. 
The transcorrelated treatment therefore modifies both the underlying particle-hole spectrum and the interaction kernel that couples these excitations. 
Consequently, the difference between conventional \ac{RPA} and \ac{TC}-\ac{RPA} cannot be interpreted as a simple additive correction to the conventional \ac{RPA} kernel. 

In this respect, \ac{TC}-\ac{RPA} shares a common conceptual structure with the Bethe-Salpeter equation formalism \cite{Blase_2020,Monsebaiz_2022} and similarity-transformed \ac{EOM} coupled-cluster methods: \cite{Nooijen_1997a,Nooijen_1997b,Nooijen_1997c} in all three cases, excitation energies are obtained from an effective eigenvalue problem in which both the underlying one-particle spectrum and the interaction between particle--hole excitations are dressed by electronic correlation.
 
\subsection{Properties of TC-RPA}

Unlike conventional \ac{RPA}, the present formalism is characterized by a non-symmetric $\bA$ block and by $\bB \neq \bC$.
The right and left eigenvectors therefore no longer have the standard Hermitian-\ac{RPA} structure, i.e.,
\begin{align}
    \bR & =
    \begin{pmatrix}
        \brX & \brY \\
        \brY & \brX
    \end{pmatrix},
    &
    \bL & =
    \begin{pmatrix}
         \brX & -\brY \\
        -\brY &  \brX
    \end{pmatrix}.
\end{align}
Instead, the right eigenvectors contain four distinct blocks,
\begin{equation}
    \bR =
    \begin{pmatrix}
        \brX_{1} & \brY_{2} \\
        \brY_{1} & \brX_{2}
    \end{pmatrix},
\end{equation}
while the corresponding left eigenvectors can be written as
\begin{equation}
    \bL =
    \begin{pmatrix}
         \brX_{2} & -\brY_{1} \\
        -\brY_{2} &  \brX_{1}
    \end{pmatrix},
\end{equation}
as shown in the \SupMat.
These blocks satisfy
\begin{subequations}
\begin{align}
	\bM \cdot \bR
	& = \bR \cdot
	\begin{pmatrix}
		\bOm & \bO \\
		\bO & -\bOm
	\end{pmatrix}, \\
	\T{\bM} \cdot \bL
	& =
	\bL \cdot
	\begin{pmatrix}
		\bOm & \bO \\
		\bO & -\bOm
	\end{pmatrix}.
\end{align}
\end{subequations}
As in conventional \ac{RPA}, the eigenvalues of the \ac{TC}-\ac{RPA} matrix occur in opposite-sign pairs $(\bOm,-\bOm)$ as a consequence of the symplectic structure of Eq.~\eqref{eq:Mdef}.

The biorthogonality condition $\T{\bL} \cdot \bR = \bI$ gives
\begin{subequations} \label{eq:biorthog}
\begin{align}
	\T{\brX_{2}} \cdot \brX_{1} - \T{\brY_{2}} \cdot \brY_{1}
	&= \bI
	= \T{\brX_{1}} \cdot \brX_{2} - \T{\brY_{1}} \cdot \brY_{2}, \\
	\T{\brX_{2}} \cdot \brY_{2} - \T{\brY_{2}} \cdot\brX_{2}
	&= \bO
	= \T{\brY_{2}} \cdot \brX_{2} - \T{\brX_{2}} \cdot \brY_{2}, \\
	-\T{\brY_{1}} \cdot \brX_{1} + \T{\brX_{1}} \cdot \brY_{1}
	&= \bO
	= -\T{\brX_{1}} \cdot \brY_{1} + \T{\brY_{1}} \cdot \brX_{1}.
\end{align}
\end{subequations}
These relations ensure the quasiboson commutation rules
\begin{align}
        \comm{ \dualQ{\mu} }{ \opQd{\nu} }  &= \delta_{\mu\nu},
        &
        \comm{ \opQd{\mu} }{ \opQd{\nu} } &= 0,
        &
        \comm{ \dualQ{\mu} }{ \dualQ{\nu} } &= 0.
\end{align}
As in conventional \ac{RPA}, a Tamm-Dancoff approximation can be obtained by setting the off-diagonal blocks to zero, i.e., $\bB=\bC=\bO$.

\subsection{RPA correlation energy}

To derive the \ac{RPA} correlation energy, we introduce a quadratic bosonic Hamiltonian $\hH_\text{B}$ associated with the \ac{RPA} eigenproblem.\cite{Holstein_1940,Beliaev_1962,Ring_2004,Ripka_1986}
As shown in the \SupMat, this Hamiltonian is non-Hermitian and reads
\begin{equation}
        \hH_\text{B}
        =
        E_{\text{TC-HF}} - \frac{1}{2} \, \Tr(\bA)
        + \frac{1}{2}
        \begin{pmatrix}
                \bb^{\dagger} & \tilde{\bb}
        \end{pmatrix}
        \cdot
        \begin{pmatrix}
                \bA & \bB \\
                \bC & \T{\bA}
        \end{pmatrix}
        \cdot
        \begin{pmatrix}
                \tilde{\bb} \\
                \bb^{\dagger}
        \end{pmatrix}.
\end{equation}
Here, $\bb^{\dagger}$ and $\tilde{\bb}$ collect the quasiboson creation and annihilation operators, which are assumed to obey the bosonic commutation relations of Eq.~\eqref{eq:qba}.

The Hamiltonian is diagonalized through a Bogoliubov-like transformation of the bosonic operators.
In the present non-Hermitian setting, this transformation involves two matrices, $\bT_1$ and $\bT_2$, satisfying
\begin{align}
	\bT_1 &= \bDel \cdot \T{\bR} \cdot \bDel,
	&
	\bT_2 &= \T{\bL},
\end{align}
where
\begin{equation}
        \bDel =
        \begin{pmatrix}
                \bI& \bO \\
                \bO & -\bm{1}
        \end{pmatrix}.
\end{equation}
After this transformation, the quadratic Hamiltonian becomes
\begin{equation}\label{eq:rpa_bog}
      \hH_\text{B} = E_{\text{TC-HF}} + E_{\text{TC-RPA}} + \sum_{\mu} \Omega_{\mu}\opQd{\mu} \dualQ{\mu},
\end{equation}
with the \ac{TC}-\ac{RPA} correlation energy
\begin{equation}\label{eq:ec_rpa_bog}
	E_{\text{TC-RPA}} = \frac{1}{2} \Tr\qty(\bOm - \bA).
\end{equation}

\subsection{Direct random-phase approximation}

The \ac{RPA} formulation derived above [see Eqs.~\eqref{eq:rpa_bog} and \eqref{eq:ec_rpa_bog}] corresponds to \ac{TC}-\ac{RPAx}, because the effective interaction entering the $\bA$, $\bB$, and $\bC$ blocks contains both direct and exchange contributions from the two- and three-body terms.
In the presence of explicit three-body interactions, however, the definition of a direct approximation is not unique.
Here, we define \ac{TC}-\ac{dRPA} by discarding exchange contributions directly in the effective interaction entering the particle-hole kernel, which gives
\begin{equation}
	U_{q_1 q_2}^{p_1 p_2}
	= V_{q_1 q_2}^{p_1 p_2} + \sum_{k} L_{k q_1 q_2}^{k p_1 p_2}.
\end{equation}
The corresponding \ac{RPA} matrix is
\begin{equation}
	\boldsymbol{\mathcal{M}}
	=
        \begin{pmatrix}
		+\bm{\mathcal{A}} & +\bm{\mathcal{B}} \\
		-\bm{\mathcal{C}} & -\T{\bm{\mathcal{A}}}
        \end{pmatrix},
\end{equation}
with
\begin{subequations}
\begin{align}
	\mathcal{A}_{ia,jb} &= \qty(\epsilon_a-\epsilon_i) \, \delta_{ab} \, \delta_{ij} + U_{i b}^{a j}, \\
	\mathcal{B}_{ia,jb} &= U_{i j}^{a b}, \\
	\mathcal{C}_{ia,jb} &= U_{a b}^{i j}.
\end{align}
\end{subequations}

\subsection{Spin adaptation}

For \ac{RHF} orbitals, the spin-orbital \ac{RPA} problem can be decomposed into singlet and triplet sectors.
This avoids the diagonalization of the full spin-orbital matrix and instead requires the diagonalization of two smaller spatial-orbital matrices, $\prescript{1}{}{\bM}$ and $\prescript{3}{}{\bM}$:
\begin{align}
	\prescript{1}{}{\bM}
	&=
	\begin{pmatrix}
		+\prescript{1}{}{\bA} & +\prescript{1}{}{\bB} \\
		-\prescript{1}{}{\bC} & -\prescript{1}{}{\T{\bA}}
        \end{pmatrix},
    &
	\prescript{3}{}{\bM}
	&=
	\begin{pmatrix}
		+\prescript{3}{}{\bA} & +\prescript{3}{}{\bB} \\
		-\prescript{3}{}{\bC} & -\prescript{3}{}{\T{\bA}}
        \end{pmatrix}.
\end{align}

For \ac{RPAx}, the singlet blocks are
\begin{subequations}
\begin{align}
        \begin{split}
        \prescript{1}{}{A}_{IA,JB}
        &= \qty(\epsilon_A - \epsilon_I) \, \delta_{IJ} \, \delta_{AB}
        + 2 \, \left\langle\left\langle \lphi{A} \, \lphi{J} | \rphi{I} \, \rphi{B} \right\rangle\right\rangle \\
	&\quad - \left\langle\left\langle \lphi{A} \, \lphi{J} | \rphi{B} \, \rphi{I} \right\rangle\right\rangle,
        \end{split} \\
        \prescript{1}{}{B}_{IA,JB}
        &= 2 \, \left\langle\left\langle \lphi{A} \, \lphi{B} | \rphi{I} \, \rphi{J} \right\rangle\right\rangle
        - \left\langle\left\langle \lphi{A} \, \lphi{B} | \rphi{J} \, \rphi{I} \right\rangle\right\rangle, \\
        \prescript{1}{}{C}_{IA,JB}
        &= 2 \, \left\langle\left\langle \lphi{I} \, \lphi{J} | \rphi{A} \, \rphi{B} \right\rangle\right\rangle
        - \left\langle\left\langle \lphi{I} \, \lphi{J} | \rphi{B} \, \rphi{A} \right\rangle\right\rangle,
\end{align}
\end{subequations}
and the triplet blocks are
\begin{subequations}
\begin{align}
	\prescript{3}{}{A}_{IA,JB} &= \qty(\epsilon_A - \epsilon_I) \, \delta_{IJ} \, \delta_{AB}
	- \left\langle\left\langle \lphi{A} \, \lphi{J} | \rphi{B} \, \rphi{I} \right\rangle\right\rangle, \\
	\prescript{3}{}{B}_{IA,JB} &= - \left\langle\left\langle \lphi{A} \, \lphi{B} | \rphi{J} \, \rphi{I} \right\rangle\right\rangle, \\
	\prescript{3}{}{C}_{IA,JB} &= - \left\langle\left\langle \lphi{I} \, \lphi{J} | \rphi{B} \, \rphi{A} \right\rangle\right\rangle.
\end{align}
\end{subequations}
The effective spatial-orbital matrix elements are defined as
\begin{equation}
	\left\langle\left\langle \lphi{P} \, \lphi{Q} | \rphi{R} \, \rphi{S} \right\rangle\right\rangle
	=
	V_{RS}^{PQ} + \sum_{k=1}^{\Ne} \, \Big[ L_{RSk}^{PQk} - L_{RkS}^{PQk} - L_{kSR}^{PQk}\Big].
\end{equation}
The \ac{RPAx} correlation energy is then obtained as\cite{Angyan_2011,Eshuis_2012}
\begin{equation}
	E_{\text{TC-RPAx}}
	=
	\frac{1}{4} \Tr\qty(\prescript{1}{}{\bOm} - \prescript{1}{}{\bA}) +
	\frac{3}{4} \Tr\qty(\prescript{3}{}{\bOm} - \prescript{3}{}{\bA}).
\end{equation}

For \ac{dRPA}, the singlet blocks are
\begin{subequations}
\begin{align}
	\prescript{1}{}{\mathcal{A}}_{IA,JB} &= \qty(\epsilon_A - \epsilon_I) \delta_{IJ} \delta_{AB} + U_{IB}^{AJ}, \\
	\prescript{1}{}{\mathcal{B}}_{IA,JB} &= 2 U_{IJ}^{AB}, \\
	\prescript{1}{}{\mathcal{C}}_{IA,JB} &= 2 U_{AB}^{IJ},
\end{align}
\end{subequations}
while the triplet blocks reduce to
\begin{subequations}
\begin{align}
	\prescript{3}{}{\mathcal{A}}_{IA,JB} &= \qty(\epsilon_A - \epsilon_I) \delta_{IJ} \delta_{AB}, \\
	\prescript{3}{}{\mathcal{B}}_{IA,JB} &= 0, \\
	\prescript{3}{}{\mathcal{C}}_{IA,JB} &= 0.
\end{align}
\end{subequations}
The triplet eigenproblem is therefore diagonal and does not contribute to the \ac{dRPA} correlation energy.
Accordingly,
\begin{equation}
	E_{\text{TC-dRPA}} = \frac{1}{2} \Tr\qty(\prescript{1}{}{\bOm} - \prescript{1}{}{\bm{\mathcal{A}}}).
\end{equation}

\section{Results}
\label{sec:res}

\subsection{Computational details}

In this work, we adopt the Jastrow factor proposed by Boys and Handy, \cite{Boys_1969} whose functional 
form and parameters are detailed in Section~II.F of Ref.~\onlinecite{Ammar_2024}.
These parameters, optimized at the single-determinant level within a variational Monte Carlo framework, were taken from the 
literature. 
Specifically, the atomic parameters were extracted from Table~V of Ref.~\onlinecite{Schmidt_1990} and the molecular 
parameters from Table~2 of Ref.~\onlinecite{Galek_2006}.

The approach to evaluating the \ac{TC} integrals is outlined in Section~II.G of Ref.~\onlinecite{Ammar_2024}.
We should mention here that the three-electron contribution to the \ac{TC}-\ac{dRPA} 
and \ac{TC}-\ac{RPAx} calculations is computed on the fly using a quadrature
grid, \cite{Becke_1988,Mura_1996,Ammar_2024} without the need to store a 6-index tensor 
(see \SupMat and Section~III.H of Ref.~\onlinecite{Ammar_2024} for more details).

All calculations were carried out with Quantum Package. \cite{Garniron_2019} 
We employed Dunning's augmented correlation-consistent basis sets, aug-cc-pV$X$Z (abbreviated as aV$X$Z), with cardinal numbers $X =$ D, T, Q, 5, and 6. \cite{Dunning_1989,Kendall_1992, Woon_1995} 
Equilibrium molecular geometries and theoretical best estimates (TBEs) for the vertical excitation energies were retrieved from the QUEST database, \cite{Loos_2018,Loos_2020_quest,Veril_2021,Loos_2025}
available at \url{https://github.com/pfloos/QUESTDB}.

\subsection{Ground-state energies}

\begin{figure}
		\includegraphics[width=\linewidth]{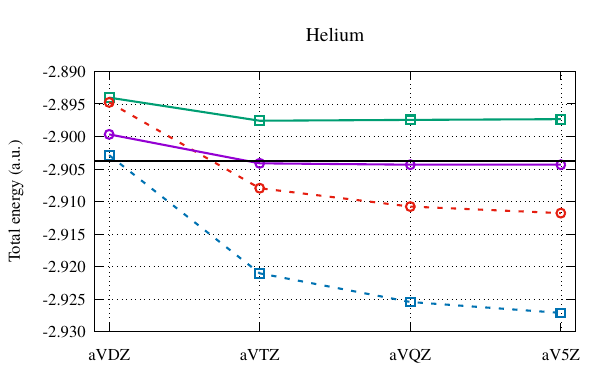}
		\includegraphics[width=\linewidth]{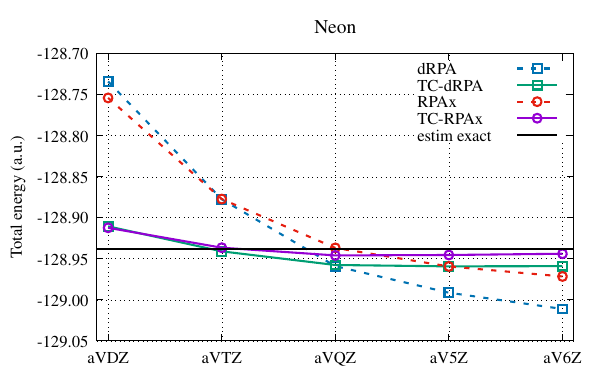}
\caption{
Basis set convergence of the total ground-state energies (in a.u.) of the He and Ne atoms.
The energies are plotted as a function of the basis set cardinal number $X$ (aug-cc-pV$X$Z) for dRPA (blue dashed), TC-dRPA (green solid), RPAx (red dashed), and TC-RPAx (purple solid).
The horizontal black line denotes the estimated exact non-relativistic ground-state energy, which serves as the \ac{CBS} reference. \cite{Oeill_2005}
Raw data can be found in the \SupMat.
}
	\label{fig:Egs_He_Ne}
\end{figure}

We begin by assessing the accuracy and basis set convergence of \ac{RPA} and \ac{TC}-\ac{RPA} ground-state energies for the helium and neon atoms.
Figure~\ref{fig:Egs_He_Ne} compares \ac{dRPA} and \ac{RPAx} with their transcorrelated counterparts, \ac{TC}-\ac{dRPA} and \ac{TC}-\ac{RPAx}, as a function of the basis set cardinal number.
Exact non-relativistic energies are also shown for reference. \cite{Oeill_2005}

Two distinct effects of the transcorrelated treatment are apparent: it substantially accelerates basis set convergence and modifies the limiting energy toward which the \ac{RPA} approximation converges. 
These effects should be distinguished from the behavior of conventional explicitly correlated F12 methods. 
By incorporating the short-range electron-electron cusp explicitly, F12 methods accelerate basis set convergence while formally preserving the \ac{CBS} limit of the underlying electronic structure method. \cite{Hattig_2012,Kong_2012}
The \ac{TC} transformation is also isospectral when treated exactly and therefore leaves the exact energy spectrum unchanged. 
In the present context, however, the \ac{RPA} approximation is applied to the transformed, non-Hermitian Hamiltonian and is not invariant under this similarity transformation. 
Consequently, conventional \ac{RPA} and \ac{TC}-\ac{RPA} need not share the same approximate \ac{CBS} limit. 
The displacement of the limiting energy observed here should therefore be attributed to the interplay between the \ac{TC} transformation and the \ac{RPA} approximation, rather than to a modification of the exact spectrum.

These two effects are particularly striking for helium.
At the \ac{dRPA} level, the error on the total energy evolves from 
$+0.8$ mH with aVDZ to $-23.4$ mH with aV5Z, 
with the energy continuing to decrease appreciably as the basis set is enlarged.
This behavior does not signal a failure of basis set convergence itself; rather, \ac{dRPA} progressively approaches its own \ac{CBS} limit, which lies substantially below the exact fermionic energy.
In contrast, \ac{TC}-\ac{dRPA} exhibits a much weaker basis set dependence, with errors ranging from 
$+9.7$ mH at aVDZ to $6.4$ mH at aV5Z 
and the energy already nearly converged at the aVTZ level.
The TC treatment therefore both accelerates convergence and profoundly alters the limiting energy.

A similar behavior is observed for \ac{RPAx}.
The conventional \ac{RPAx} energy continues to evolve significantly with increasing basis set size, with its error changing from 
$+8.9$ mH at aVDZ to $-8.1$ mH
at aV5Z as it approaches its own \ac{CBS} limit.
By contrast, \ac{TC}-\ac{RPAx} converges much more rapidly: its energy is essentially stabilized from aVTZ onward and remains remarkably close to the exact non-relativistic value of $-2.9037$ Hartree.
The corresponding error decreases from 
$+4.0$ mH at aVDZ to only $-0.6$ mH at aV5Z.

The same qualitative picture emerges for neon.
The conventional \ac{dRPA} energy retains a pronounced basis set dependence throughout the sequence, with the error evolving from 
$+203.7$ mH at aVDZ to $-73.6$ mH at aV6Z.
\ac{TC}-\ac{dRPA} converges considerably faster, becoming nearly stationary at the largest basis sets and remaining close to the exact energy, 
with errors ranging from $+27.2$ mH to $-21.4$ mH.
Likewise, the \ac{RPAx} error evolves 
from $+183.7$ mH at aVDZ to $-33.9$ mH at aV6Z, 
whereas \ac{TC}-\ac{RPAx} displays a much weaker basis set dependence and rapidly approaches a limiting energy very close to the exact reference, with errors 
decreasing from $+25.3$ mH to only $-6.3$ mH.

Overall, Fig.~\ref{fig:Egs_He_Ne} highlights two complementary benefits of the transcorrelated treatment.
First, both \ac{TC}-\ac{dRPA} and \ac{TC}-\ac{RPAx} exhibit substantially faster convergence with respect to the one-electron basis, reaching near-asymptotic energies with considerably smaller basis sets than their conventional counterparts.
Second, the TC transformation modifies the effective many-body space explored by the \ac{RPA} approximation, leading to \ac{CBS} limits that lie much closer to the exact fermionic ground-state energies.
These results suggest that the transcorrelated formulation simultaneously alleviates the slow basis set convergence of conventional \ac{RPA} and reduces the energetic consequences of the unphysical bosonic components introduced by the quasiboson approximation.


%
%
%
%
%
%
%
%
%
%

\begin{figure*}
		\includegraphics[width=0.49\linewidth]{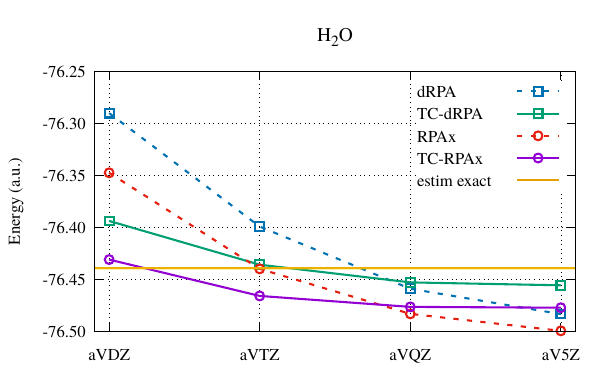}
		\includegraphics[width=0.49\linewidth]{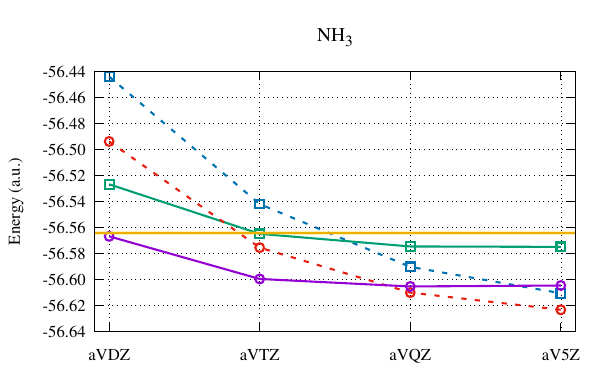}
		\includegraphics[width=0.49\linewidth]{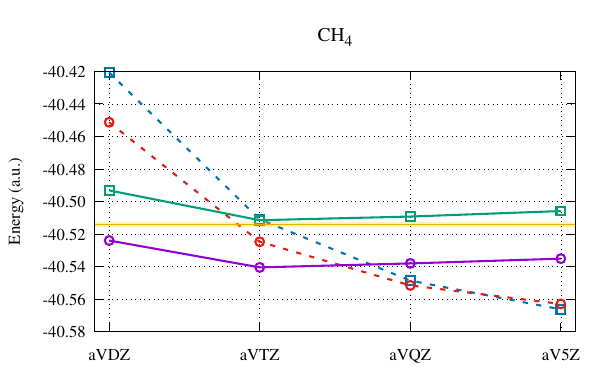}
		\includegraphics[width=0.49\linewidth]{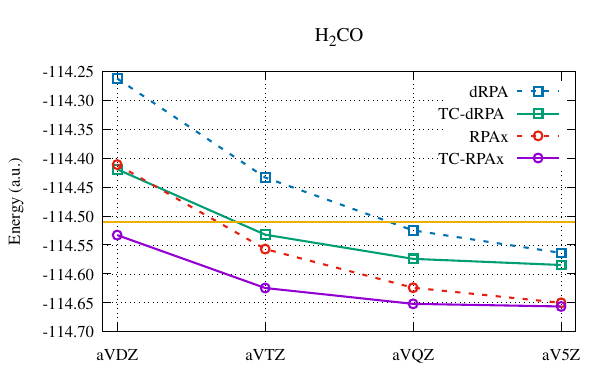}
\caption{
Basis-set convergence of the total ground-state energies (in a.u.) of \ce{H2O}, \ce{NH3}, \ce{CH4}, and \ce{H2CO} as a function of the basis set cardinal number $X$ (aug-cc-pV$X$Z).
Results are shown for dRPA (blue dashed), TC-dRPA (green solid), RPAx (red dashed), and TC-RPAx (purple solid).
The horizontal yellow line represents the estimated exact non-relativistic ground-state energy. \cite{Galek_2006,Caffarel_2016}
Raw data can be found in the \SupMat.
}
	\label{fig:E_molec}
\end{figure*}

We now extend our analysis to the ground-state energies of a set of molecules comprising \ce{H2O}, \ce{NH3}, \ce{CH4}, and \ce{H2CO}.
Figure~\ref{fig:E_molec} compares the total energies obtained with conventional \ac{RPA} and \ac{TC}-\ac{RPA} as a function of the basis set cardinal number.
Estimated exact non-relativistic ground-state energies are also shown for reference (Ref.~\onlinecite{Caffarel_2016} for \ce{H2O} and Ref.~\onlinecite{Galek_2006} for \ce{NH3}, 
\ce{CH4}, and \ce{H2CO}).

The molecular results confirm the trends observed for He and Ne.
Most notably, the \ac{TC}-\ac{RPA} variants converge significantly faster with respect to the one-electron basis than their conventional counterparts, for both \ac{dRPA} and \ac{RPAx}.
This acceleration is particularly pronounced for \ac{TC}-\ac{RPAx}, whose energies are already close to their limiting values at the aVTZ or aVQZ level for most systems.
\ac{TC}-\ac{dRPA} exhibits a similarly reduced basis set dependence, although the convergence is somewhat more system dependent, especially for \ce{H2CO}.
In contrast, the conventional \ac{dRPA} and \ac{RPAx} energies continue to vary appreciably up to the largest basis sets considered.

Besides accelerating basis set convergence, the TC treatment also modifies the limiting energies reached by the \ac{RPA} approximations.
At the aV5Z level, the absolute errors of \ac{dRPA} are 
$44.2$ mH, $46.3$ mH, $52.1$ mH, and $53.9$ mH 
for \ce{H2O}, \ce{NH3}, \ce{CH4}, and \ce{H2CO}, respectively, compared with 
$16.5$ mH, $10.6$ mH, $8.4$ mH, and $74.7$ mH 
for \ac{TC}-\ac{dRPA}.
The improvement is therefore particularly systematic for \ac{TC}-\ac{dRPA}, which yields remarkably accurate energies for \ce{H2O}, \ce{NH3}, and \ce{CH4}.
For \ac{RPAx}, the corresponding errors decrease from 
$60.2$ mH, $59.0$ mH, $48.7$ mH, and $140.1$ mH to $38.0$ mH, $40.3$ mH, $20.9$ mH, and $146.3$ mH 
upon introducing the TC treatment.
Thus, while \ac{TC}-\ac{RPAx} substantially improves the energies of \ce{H2O}, \ce{NH3}, and \ce{CH4}, no 
improvement in absolute accuracy is observed for \ce{H2CO} at the aV5Z level.

These comparisons should, however, be interpreted in light of the markedly different basis set convergence rates of the conventional and \ac{TC} approaches.
At aV5Z, the \ac{TC}-\ac{RPA} energies are already close to their respective \ac{CBS} limits, whereas the conventional \ac{RPA} energies, particularly those of \ac{dRPA}, still exhibit a noticeable basis set dependence.
Consequently, similar errors at a finite basis set level do not necessarily imply similar accuracies at the \ac{CBS} limit.
As discussed above, the conventional and transcorrelated \ac{RPA} formulations effectively explore different approximate many-body spaces and therefore need not converge toward the same \ac{CBS} energy.

Overall, the molecular results reinforce the two main effects of the transcorrelated treatment already identified for the atomic systems.
First, transcorrelation substantially accelerates basis set convergence for both \ac{dRPA} and \ac{RPAx}.
Second, it modifies the corresponding \ac{CBS} limits, often bringing them closer to the exact fermionic ground-state energies, although this improvement in absolute accuracy is system- and approximation-dependent.
Importantly, these two effects are distinct: faster basis set convergence does not necessarily imply a more accurate \ac{CBS} limit.
Among the systems considered here, the improvement in accuracy is particularly systematic for \ac{TC}-\ac{dRPA}, whereas \ac{TC}-\ac{RPAx} exhibits a more system-dependent accuracy despite its consistently faster basis set convergence.

%
%

\begin{table}[t]
\caption{
Vertical excitation energies (in eV) of the two lowest Rydberg states of \ce{H2O}, computed with dRPA, TC-dRPA, RPAx, and TC-RPAx using the aug-cc-pV$X$Z basis sets.
The theoretical best estimates (TBEs) at the CBS limit are taken from Ref.~\onlinecite{Loos_2025}.
}
\label{tab:H2O_excitations}
\begin{ruledtabular}
\begin{tabular}{lcccc}
	& \multicolumn{4}{c}{Cardinal number $X$} \\
	\cline{2-5}
	Method & D & T & Q & 5 \\
	\hline
	\multicolumn{5}{c}{
		$1\,^{1}\mathrm{A_1} \rightarrow 1\,^{1}\mathrm{B_1}$
		$\left(n\rightarrow3s\right)$,
		TBE $=7.672$
	} \\
	dRPA    & 14.909 & 14.752 & 14.680 & 14.581 \\
	TC-dRPA & 14.601 & 14.423 & 14.337 & 14.230 \\
	RPAx    &  8.625 &  8.640 &  8.639 &  8.639 \\
	TC-RPAx &  8.638 &  8.640 &  8.639 &  8.639 \\
	\hline
	\multicolumn{5}{c}{
		$1\,^{1}\mathrm{A_1} \rightarrow 1\,^{1}\mathrm{A_2}$
		$\left(n\rightarrow3p\right)$,
		TBE $=9.449$
	} \\
	dRPA    & 15.455 & 15.194 & 15.070 & 14.900 \\
	TC-dRPA & 15.122 & 14.832 & 14.687 & 14.500 \\
	RPAx    & 10.306 & 10.312 & 10.308 & 10.306 \\
	TC-RPAx & 10.345 & 10.340 & 10.333 & 10.333 \\
\end{tabular}
\end{ruledtabular}
\end{table}

\begin{table}[t]
\caption{
Vertical excitation energies (in eV) of the two lowest Rydberg states of \ce{NH3}, computed with dRPA, TC-dRPA, RPAx, and TC-RPAx using the aug-cc-pV$X$Z basis sets.
The theoretical best estimates (TBEs) at the CBS limit are taken from Ref.~\onlinecite{Loos_2025}.
}
\label{tab:NH3_excitations}
\begin{ruledtabular}
\begin{tabular}{lcccc}
	& \multicolumn{4}{c}{Cardinal number $X$} \\
	\cline{2-5}
	Method & D & T & Q & 5 \\
	\hline
	\multicolumn{5}{c}{
		$1\,^{1}\mathrm{A_1} \rightarrow 1\,^{1}\mathrm{A_2}$
		$\left(n\rightarrow3s\right)$,
		TBE $=6.627$
	} \\
	dRPA    & 12.734 & 12.559 & 12.476 & 12.398 \\
	TC-dRPA & 12.704 & 12.542 & 12.453 & 12.380 \\
	RPAx    &  7.414 &  7.423 &  7.427 &  7.428 \\
	TC-RPAx &  7.546 &  7.558 &  7.554 &  7.556 \\
	\hline
	\multicolumn{5}{c}{
		$1\,^{1}\mathrm{A_1} \rightarrow 1\,^{1}\mathrm{E}$
		$\left(n\rightarrow3p\right)$,
		TBE $=8.190$
	} \\
	dRPA    & 13.323 & 13.028 & 12.896 & 12.743 \\
	TC-dRPA & 13.314 & 13.032 & 12.894 & 12.745 \\
	RPAx    &  8.888 &  8.877 &  8.875 &  8.869 \\
	TC-RPAx &  9.096 &  9.088 &  9.078 &  9.074 \\
\end{tabular}
\end{ruledtabular}
\end{table}

  
\begin{table}[t]
\caption{
Vertical excitation energies (in eV) of the lowest valence and Rydberg states of \ce{H2CO}, computed with dRPA, TC-dRPA, RPAx, and TC-RPAx using the aug-cc-pV$X$Z basis sets.
The theoretical best estimates (TBEs) at the CBS limit are taken from Ref.~\onlinecite{Loos_2025}.
}
\label{tab:H2CO_excitations}
\begin{ruledtabular}
\begin{tabular}{lcccc}
	& \multicolumn{4}{c}{Cardinal number $X$} \\
	\cline{2-5}
	Method & D & T & Q & 5 \\
	\hline
	\multicolumn{5}{c}{
		$1\,^{1}\mathrm{A_1} \rightarrow 1\,^{1}\mathrm{A_2}$
		$\left(n\rightarrow\pi^\star\right)$,
		TBE $=3.969$
	} \\
	dRPA    & 12.988 & 12.831 & 12.754 & 12.689 \\
	TC-dRPA & 12.288 & 12.114 & 11.999 & 11.911 \\
	RPAx    &  4.379 &  4.396 &  4.395 &  4.394 \\
	TC-RPAx &  4.250 &  4.285 &  4.281 &  4.280 \\
	\hline
	\multicolumn{5}{c}{
		$1\,^{1}\mathrm{A_1} \rightarrow 1\,^{1}\mathrm{B_2}$
		$\left(n\rightarrow3s\right)$,
		TBE $=7.269$
	} \\
	dRPA    & 13.473 & 13.247 & 13.145 & 13.003 \\
	TC-dRPA & 12.710 & 12.460 & 12.318 & 12.143 \\
	RPAx    &  8.566 &  8.588 &  8.588 &  8.586 \\
	TC-RPAx &  8.220 &  8.248 &  8.229 &  8.221 \\
\end{tabular}
\end{ruledtabular}
\end{table}

\subsection{Excited-state energies}

We now turn to vertical excitation energies and investigate whether the benefits of the 
transcorrelated treatment observed for ground-state energies carry over to excited states.
Table~\ref{tab:H2O_excitations} compares \ac{dRPA} and \ac{RPAx} with their transcorrelated 
counterparts for two vertical Rydberg excitations of \ce{H2O}.
In contrast to the ground-state energies, the TC treatment does not lead to a systematic 
improvement of the excitation energies.

For the $n \rightarrow 3s$ transition (TBE = 7.672~eV), \ac{dRPA} strongly overestimates 
the excitation energy, yielding 14.581~eV at the aV5Z level.
The TC treatment lowers this value by approximately $0.3$--$0.4$~eV across the basis set 
sequence, reaching 14.230~eV at aV5Z.
Although this shift is non-negligible, it is far too small to remedy the large intrinsic 
error of \ac{dRPA} for this transition.
\ac{RPAx} provides a much more accurate description, with an aV5Z excitation energy of 
8.639~eV, but the TC transformation has essentially no effect: \ac{RPAx} and \ac{TC}-\ac{RPAx} are 
virtually indistinguishable beyond aVDZ.

A similar picture emerges for the $n \rightarrow 3p$ transition (TBE = 9.449~eV).
The TC transformation lowers the \ac{dRPA} excitation energy by approximately 
$0.3$--$0.4$~eV, from 14.900~eV to 14.500~eV at the aV5Z level, but the resulting 
\ac{TC}-\ac{dRPA} value remains more than 5~eV above the reference.
For \ac{RPAx}, the effect of the TC treatment is again very small, with changes of only a 
few hundredths of an eV across the entire basis set sequence.
Interestingly, in this case the TC correction slightly increases the \ac{RPAx} excitation 
energy and therefore does not improve agreement with the TBE.

The analysis is extended to \ce{NH3} and \ce{H2CO} in 
Tables~\ref{tab:NH3_excitations} and \ref{tab:H2CO_excitations}, respectively.
These results reinforce the absence of a systematic effect of the TC transformation 
on vertical excitation energies, while also revealing a pronounced dependence on the 
system and the nature of the transition.

For \ce{NH3}, the TC correction has an almost negligible effect on \ac{dRPA}.
At the aV5Z level, the excitation energies change by only $-0.018$~eV for the 
$n \rightarrow 3s$ transition and $+0.002$~eV for the $n \rightarrow 3p$ transition.
Consequently, the very large \ac{dRPA} errors remain essentially unchanged, with \ac{TC}-\ac{dRPA} 
overestimating the corresponding TBEs by approximately 5.7 and 4.5~eV, respectively.
The effect is somewhat larger for \ac{RPAx}, for which the TC transformation increases the 
aV5Z excitation energies by 0.128~eV for the $n \rightarrow 3s$ transition and 
0.205~eV for the $n \rightarrow 3p$ transition.
Since the conventional \ac{RPAx} values already lie above the corresponding TBEs, these 
shifts slightly deteriorate the agreement with the reference values.

The behavior of \ce{H2CO} is more contrasted.
For the valence $n \rightarrow \pi^\star$ transition (TBE = 3.969~eV), the TC 
transformation produces a substantial reduction of the \ac{dRPA} excitation energy, from 
12.689~eV to 11.911~eV at the aV5Z level.
Despite this sizable $0.778$~eV shift, \ac{TC}-\ac{dRPA} still strongly overestimates the 
reference value, demonstrating that the TC transformation alone cannot remedy the 
intrinsic limitations of the \ac{dRPA} response kernel.
For \ac{RPAx}, the corresponding shift is smaller but favorable, lowering the excitation 
energy from 4.394~eV to 4.280~eV and thereby improving agreement with the TBE.

For the Rydberg $n \rightarrow 3s$ transition of \ce{H2CO} (TBE = 7.269~eV), the TC 
transformation also produces appreciable shifts.
At the aV5Z level, it lowers the \ac{dRPA} excitation energy by 0.860~eV, from 13.003~eV 
to 12.143~eV, while the \ac{RPAx} excitation energy decreases by 0.365~eV, from 8.586~eV 
to 8.221~eV.
The latter represents a noticeable improvement over conventional \ac{RPAx}, although the 
\ac{TC}-\ac{RPAx} value remains approximately 0.9~eV above the reference.

An important aspect of these results is that most of the excited states considered here 
are Rydberg states.
Their accurate description requires substantial radial flexibility in the one-electron 
basis and, in particular, the presence of sufficiently diffuse basis functions.
The transcorrelated approach employed here is not specifically designed to address this 
source of basis set incompleteness.
Rather, by explicitly incorporating short-range electron correlation through the Jastrow 
factor, it primarily alleviates the need for the high-angular-momentum components required 
to describe the electron-electron cusp.
Consequently, one should not necessarily expect the TC transformation to eliminate the 
slow convergence associated with the radial and diffuse character of Rydberg states, 
and augmented basis sets remain essential for their accurate description.

In this respect, the valence $n \rightarrow \pi^\star$ excitation of \ce{H2CO} is 
particularly noteworthy.
Unlike the predominantly Rydberg transitions considered above, this excitation does not 
rely as strongly on the diffuse radial part of the basis.
The more pronounced effect of the \ac{TC} transformation observed for this transition may 
therefore indicate that valence excitations benefit differently from the improved 
description of short-range correlation.
However, the present dataset contains too few valence excitations to establish a general 
trend, and a broader investigation would be required to substantiate this interpretation.

Overall, these results reveal a clear contrast between ground-state and excitation 
energies.
For ground-state energies, the transcorrelated treatment systematically accelerates 
basis set convergence and substantially modifies the corresponding CBS limits.
For vertical excitation energies, however, no similarly systematic improvement is 
observed.
The magnitude and even the sign of the \ac{TC}-induced shifts depend strongly on the system, 
the transition, and the underlying \ac{RPA} approximation.
In particular, the \ac{TC} transformation does not resolve the severe overestimation of 
excitation energies obtained with \ac{dRPA}, indicating that the dominant deficiency lies 
in the approximate response kernel itself.
The inclusion of exchange through \ac{RPAx} is far more consequential, reducing errors by 
several eV in many cases before the \ac{TC} transformation is even introduced.
Once exchange is included, the additional effect of \ac{TC} ranges from essentially negligible 
for \ce{H2O}, to slightly detrimental for \ce{NH3}, and beneficial for the transitions 
considered in \ce{H2CO}.

Several factors may contribute to this contrasting behavior.
First, excitation energies are energy differences, so that substantial TC-induced changes 
in the absolute ground- and excited-state energies may partially cancel.
Second, the Jastrow factor employed in the present calculations is optimized exclusively 
for the ground state and therefore does not explicitly account for the correlation 
characteristics of the target excited states.
Finally, for the predominantly Rydberg states considered here, the dominant basis set 
requirement is associated with radial diffuseness rather than the short-range correlation 
effects that the TC transformation is primarily designed to capture.
These observations suggest several possible routes for further improvement, including 
state-specific or state-averaged optimization of the Jastrow factor and the development 
of improved response kernels tailored to the transcorrelated Hamiltonian.
A more systematic investigation of valence excitations would also be particularly 
valuable to disentangle the respective roles of short-range correlation and radial 
basis set incompleteness.

\section{Conclusion}
\label{sec:ccl}

In this work, we have derived the \ac{RPA} equations for a non-Hermitian Hamiltonian 
containing up to three-body interactions, with particular emphasis on the 
\ac{TC} Hamiltonian.
This formulation generalizes the standard \ac{RPA} framework to Jastrow-correlated 
wave functions and provides a rigorous foundation for computing both ground-state 
correlation energies and vertical excitation energies within \ac{TC}-\ac{RPA}.

For ground-state energies, our results reveal two major benefits of the 
transcorrelated treatment.
First, both \ac{TC}-\ac{dRPA} and \ac{TC}-\ac{RPAx} exhibit substantially faster basis set convergence 
than their conventional counterparts, reaching near-asymptotic energies with 
considerably smaller one-electron basis sets.
Second, the \ac{TC} transformation modifies the limiting energies reached by the \ac{RPA} 
approximations, often bringing them significantly closer to the exact 
non-relativistic ground-state energies.
These two effects are distinct: faster basis set convergence does not necessarily 
imply a more accurate \ac{CBS} limit.
Nevertheless, across the atoms and small molecules considered here, the \ac{TC} treatment 
provides a systematic and substantial improvement in basis set convergence, together 
with generally improved ground-state energetics.
These findings demonstrate the effectiveness of explicitly incorporating short-range 
correlation through the Jastrow factor within the \ac{RPA} framework.

The picture is more nuanced for vertical excitation energies.
Unlike for ground-state energies, the \ac{TC} transformation does not produce a systematic 
improvement across the systems and transitions considered.
The magnitude and even the sign of the \ac{TC}-induced shifts depend on the system, the 
nature of the excitation, and the underlying \ac{RPA} approximation.
In particular, \ac{TC} does not remedy the severe overestimation of excitation energies 
obtained with \ac{dRPA}, indicating that the dominant deficiency lies in the approximate 
response kernel itself.
The inclusion of exchange through \ac{RPAx} has a much larger impact on the excitation 
energies, while the additional effect of the \ac{TC} transformation is strongly 
system dependent.

Several factors may account for this contrasting behavior.
The Jastrow factor employed in the present calculations is optimized exclusively 
for the ground state and therefore does not explicitly account for the specific 
correlation characteristics of the target excited states.
Moreover, most of the excitations considered here are Rydberg in nature and require 
substantial radial flexibility through diffuse basis functions.
By explicitly describing short-range electron correlation, the TC transformation 
primarily alleviates basis set incompleteness associated with the angular description 
of the electron-electron cusp, rather than the radial incompleteness that is critical 
for Rydberg states.
Interestingly, the results obtained for the valence $n \rightarrow \pi^\star$ 
excitation of \ce{H2CO} suggest that valence excitations may respond differently to 
the \ac{TC} treatment, although a broader set of transitions is required to establish 
whether this represents a general trend.

Overall, this work establishes a general \ac{RPA} framework for non-Hermitian Hamiltonians 
with three-body interactions and demonstrates the considerable potential of its 
application to transcorrelated Hamiltonians.
For ground-state energetics, \ac{TC}-\ac{RPA} combines substantially accelerated basis set 
convergence with generally improved accuracy, making it a promising route toward 
compact and efficient correlated calculations.
For excited states, the present results highlight a more complex interplay between 
the Jastrow factor, the response kernel, and the radial and angular requirements of 
the one-electron basis.
They therefore point toward several directions for future developments, including 
improved response kernels tailored to the \ac{TC} Hamiltonian, state-specific or 
state-averaged optimization of the Jastrow factor, and a systematic investigation 
of valence excitations.

\begin{acknowledgments}
This work was performed using HPC resources from GENCI-TGCC
(gen1738,gen12363) and from CALMIP (Toulouse) under allocation
2026-18005, and was also supported by the European Centre of
Excellence in Exascale Computing TREX --- Targeting Real Chemical
Accuracy at the Exascale. This project has received funding from the
European Union's Horizon 2020 --- Research and Innovation program ---
under grant agreement no.~952165.
A.A., A.S., and P.F.L.~also acknowledge funding from the European Research Council (ERC) under the European Union’s Horizon 2020 research and innovation programme (Grant agreement No.~863481).
\end{acknowledgments}

\section*{Supplementary Material}
See the \SupMat for additional theoretical details for the derivation and implementation of the \ac{TC}-\ac{RPA} formalism, as well as the raw data associated with the figures. 

\section*{Data availability statement}
The data that supports the findings of this study are available within the article and its supplementary material.

\section*{References}
\bibliography{main}

\end{document}